\documentclass[10pt,reqno]{amsart}
\usepackage[latin9]{inputenc}
\usepackage{amsmath}
\usepackage{amssymb}
\usepackage{babel}
\usepackage{comment}
\usepackage{amssymb,mathrsfs,color}
\usepackage{pinlabel}
\usepackage{amsmath, amsfonts, amsthm, verbatim, amssymb}
\usepackage{epstopdf, mathtools}
\usepackage{cite}
\usepackage{fullpage}
\newcommand{\bs}{\boldsymbol}
\newcommand{\cl}{\mathcal}
\allowdisplaybreaks[3]
\numberwithin{equation}{section}

\newtheorem{thm}{Theorem}[section]
\begin{document}
\title{Balance laws versus the Principle of Virtual Work and the limited scope of Noll's theorem}
\author{C. Rodriguez and F. dell'Isola}

\begin{abstract}
The relationship between balance laws and the Principle of Virtual Work as well as the structure of contact interactions in continua remain foundational issues in Mechanics. In this work, we revisit these issues within the distributional framework emphasized by Paul Germain. We show that while the Principle of Virtual Work implies balance of forces and moments for $n$th-gradient continua, balance laws alone do not suffice to characterize equilibrium for $n \geq 2$. We then reexamine Noll's classical theorem asserting that surface contact forces depend solely on the unit normal to the surface and identify the precise role of his additional assumptions, namely the absence of edge and wedge contact interactions and the boundedness of the surface contact density on the space of oriented surfaces. We demonstrate that these hypotheses fail for general higher-gradient continua. Consequently, the presence of curvature-dependent surface contact forces in such materials does not conflict with Noll's theorem and refutes his claim of their nonexistence.
\end{abstract}

\maketitle

\section{Introduction}

From its earliest foundations, two central postulations have been considered in Mechanics: the Principle of Virtual Work, and balance of forces and moments. Often attributed to Archytas of Tarentum \cite{spagnuolo2025early}, the Principle of Virtual Work may be regarded as the more fundamental principle in Mechanics, since all of the balance laws of Mechanics can be derived from it. This viewpoint was forcefully advanced by d'Alembert and Lagrange, who developed a unified analytical framework for finite dimensional constrained systems and for hydrodynamics. In the final note of the last edition of Lagrange's \textit{M\'ecanique Analytique}, the first formulation of the Principle of Virtual Work applicable to general continua was presented. This formulation was subsequently taken up and significantly extended by Piola, who developed a theory of continua whose internal interactions depend on gradients of deformation of arbitrary order \cite{dellIsola2014Piola1, dellIsola2014Piola2}. Piola also proved that, for the most general class of continua in which the placement is an admissible kinematical variable, equilibrium necessarily requires the balance of forces and the balance of moments of forces, a result later referred to as ``Piola's theorem" by Truesdell \cite{truesdell_tupin1960}. 

Higher-gradient continuum theories have become necessary for accurately modeling the unusual mechanics of pantographic sheets and related metamaterials. In woven pantographic fabrics, consisting of two orthogonal fiber networks with pivot connections, homogenization leads to a second-gradient elastic surface exhibiting anisotropy and even subcoercive energy characterized by vanishing stiffness under uniform modes. Such models successfully reproduce experimentally observed uncommon behaviors \cite{Giorgio2017Pantographic,Ciallella2023Spira,Turco2022Hencky,ErdenYildizdag2023Damage}. Likewise, detailed mesoscale models of pantographic cells demonstrate that hinge compliance produces dramatically different strain-energy distributions under compression versus tension \cite{Spagnuolo2021Mesoscale,Sanna2025Hinge}, a nonlocal and curvature-sensitive effect that cannot be captured by a classical first-gradient continuum.

One- and three-dimensional pantographic architectures further provide compelling evidence for both the necessity and the predictive power of higher-gradient continuum theories. The work \cite{Barchiesi2024MicroPantograph} reports that three-dimensional micro-scale pantographs combine the in-plane resilience of two-dimensional pantographs with additional features such as pronounced softening in low loading regimes, phenomena that fall outside the predictive capacity of first-gradient models. Similarly, engineered bar-and-hinge lattices such as the recently proposed ZAPAB metamaterials exhibit peculiar floppy modes and bending responses governed by curvature gradients. The works \cite{Moschini2025ZAPAB, Murcia2025computational} show that a one-dimensional continuum with energy depending solely on the spatial derivative of curvature is required to reproduce ZAPAB simulations. The work \cite{Rizzi2026ThirdGradient} design a Hart's-antiparallelogram lattice whose homogenized limit is a third-gradient rod, with elastic energy vanishing under constant-curvature deformations so that only the spatial derivative of curvature contributes. More generally, consistency and positive definiteness in discrete-to-continuum homogenization often necessitate generalized gradient theories, since Cauchy models may be incompatible with the underlying microstructure \cite{Eremeyev2026Positive}. Even duoskelion beam arrays, exhibiting axial-flexural coupling, require enriched descriptions \cite{Barchiesi2024Duoskelion}.

Variational formulations are particularly well suited to the modeling of pantographic sheets and ZAPAB-type metamaterials, since their intrinsically higher-gradient mechanical response manifested through curvature-dependent stiffness, nonlocal interactions, and edge forces, can be incorporated by postulating an energy functional whose first variation yields both the bulk equilibrium equations and the associated higher-order boundary interactions \cite{Greco2020Kirchhoff,Baroudi2019Elastica,Barretta2025Trusses,Giorgio2020DiscreteRod}. These formulations are directly amenable to numerical implementation and can be extended to account for dissipation through Hamilton-Rayleigh structures \cite{Ciallella2022Rayleigh,Aretusi2024Timoshenko}. 

Taken together, these examples show that curvature-dependent contact surface interactions, nonlocal coupling across microstructural cells, and wedge and edge contact forces are intrinsic to these metamaterials, and that higher-gradient continua provide the natural framework for capturing their multi-scale elastic response \cite{Barchiesi2019State,Fedele2022Coupling,Barchiesi2019Homogenization,Desmorat2025Micromorphic}.

However, the conclusion of Noll's theorem, see \cite{Noll1974FoundationsReprint, Truesdell1991FirstCourseRCMVol1}, asserts that the surface contact force density associated with a system of forces depends only on position, time, and the unit normal to the oriented surface over which it is integrated to produce the resultant surface contact force, see the following Section~4 for a precise formulation. Immediately following the statement of the theorem, Noll comments on the scope of this conclusion in a footnote on page~41 of \cite{Noll1974FoundationsReprint}, where he writes:
\begin{quotation}
	The assertion of this theorem appears in all of the past literature as an assumption. It has been proposed occasionally that one should weaken this assumption and allow the stress to depend not only on the tangent plane at $\boldsymbol x$, but also on the curvature of the surface $c$ at $\boldsymbol x$. The theorem given here shows that such dependence on the curvature, or on any local property of the surface at $\boldsymbol x$, is impossible.
\end{quotation}
This statement has led some researchers to conclude that higher-gradient continua, in which surface contact forces may depend on curvature, must contain an internal inconsistency. 

In this paper, we demonstrate the following.
\begin{itemize}
	\item Within the distributional framework formulated in Section~2, the Principle of Virtual Work implies balance of forces and moments for $n$th-gradient continua. However, the converse implication holds only for first-gradient continua, that is, continua whose internal work functional depends solely on first-order gradients of the placement. In higher-gradient theories, balance laws alone do not suffice to characterize equilibrium configurations.
	
	\item We show that the conclusion of Noll's theorem rests essentially on implicit additional assumptions, namely the absence of wedge and edge interactions and the boundedness of the surface contact force density on the space of oriented surfaces. Since these hypotheses are not satisfied by general higher-gradient continua, the curvature-dependent surface forces and edge interactions arising in such theories lie outside the scope of the theorem and do not signal any logical inconsistency.
\end{itemize}

Our exposition makes systematic use of the Schwartz theory of distributions, a framework whose importance in Continuum Mechanics has long been underappreciated, despite Paul Germain's sustained emphasis on its foundational character \cite{germain2020memocs, germain1973microstructure}. The essential point may be stated succinctly. The internal work functional must be linear and continuous with respect to a class of admissible virtual displacements containing smooth vector fields defined on the current configuration $\cl B$ of the body, the closure of a bounded domain in three dimensional Euclidean space. Therefore, internal work functionals are distributions supported in the body\footnote{In this work, if $\cl B$ is the closure of a bounded domain in three-dimensional Euclidean space, the topological dual $C^\infty(\cl B)'$ of the Fr\'echet space $C^\infty(\cl B)$ (endowed with its standard Fr\'echet topology) is denoted by $\cl E'(\cl B)$, the space of distributions whose support is contained in $\cl B$.}. This requirement forms a core component of the postulation introduced by d'Alembert and subsequently formalized by Lagrange and Piola. Although the tools of modern functional analysis were not yet available to these authors, Lagrange and Piola nevertheless exploited the analytical methods at their disposal with remarkable effectiveness, thereby steering the development of Mechanics toward a mathematically coherent and enduring formulation. These postulations adopted by Lagrange and Piola imply that the theory of distributions furnishes a sufficiently general analytical framework for a broad class of problems in Continuum Mechanics\footnote{Beyond the metamaterial examples discussed above, the Principle of Virtual Work naturally accommodates energetic boundary surfaces arising from surface tension or more general surface-substrate interactions \cite{GurtinMurd75,SteigOgden99,rodriguez2024elastic}, and extends seamlessly to fracture models \cite{Gorbushinetal20,rodriguez2024elastic,Rodriguez2024b,Zem17StraightCrack,Zem20Multiple,Zem21Penny,RodriguezZem26}. Admitting smooth vector fields as virtual displacements does not mean that all admissible variations are smooth, and thus does not preclude crack surfaces or other singular interfaces.}. 

A celebrated theorem of Laurent Schwartz \cite{Schwartz1950Distributions} states that for every distribution $\boldsymbol F$ supported in $\cl B$, there exist $n \in \mathbb N \cup \{0\}$  and type $(i,1)$ tensor valued measures $d\boldsymbol T_i$ for $i = 0, \ldots, n$, supported in $\mathcal B$, such that for every $\boldsymbol \varphi \in C^{\infty}(\mathcal B)$ one has\footnote{Throughout this work, we employ the Einstein summation convention, and we identify the translation space of three-dimensional Eucidean space with $\mathbb R^3$ by fixing an orthonormal basis $\{\bs e_j = \bs e^j \}_{j = 1}^3$. The contraction of a tensor valued measure of type $(i,1)$, $d \boldsymbol T = dT_k^{j_1\ldots j_i} \bs e^k \otimes \bs e_{j_1} \otimes \bs e_{j_i}$ with $\nabla^i \bs \varphi$ is denoted by $\nabla^i \bs \varphi \cdot d\boldsymbol T := \varphi^k_{,j_1\ldots j_i} dT_k^{j_1\ldots j_i}$.}
\begin{align}
\langle \boldsymbol F, \boldsymbol \varphi \rangle
=
\sum_{i=0}^{n} \int_{\mathcal B} \nabla^i \bs \varphi \cdot d \bs T_i. \label{eq:schwartz}
\end{align}
An internal work functional given by the right-hand side of \eqref{eq:schwartz} is of order $n$ and models a $n$th-gradient continuum  when
$d\boldsymbol T_{n}$ is not identically zero. In particular, we note that:
\begin{itemize}
\item When the deformation energy is written as an integral with respect to Lebesgue measure with density function $W$ that depends on the gradients of the placement up to order $n$, its first variation yields the internal work functional, which is itself of order $n$.
\item As done by Lagrange, Piola and then Cauchy, it is natural
to start the analysis with first-gradient continua.
\item Objectivity, or material frame indifference, requires that for an internal work functional we must have $d\boldsymbol T_{0}=0.$ 
\end{itemize}
This representation reveals the natural emergence of potential higher order contact interactions, including surface, wedge and edge forces, as intrinsic features of internal work functionals when one regards them as distributions.

\subsection{Outline} 
In Section 2, we formulate the Principle of Virtual Work within the functional analytic framework introduced by Paul Germain. Internal and external interactions are modeled via distributions acting on admissible virtual displacements, and we briefly discuss how internal work functionals constrain the class of admissible external work functionals.

In Section 3, we investigate the relationship between balance laws and the Principle of Virtual Work. We first reestablish, within our functional analytic setting, Piola's result that for $n$th-gradient continua the Principle of Virtual Work implies balance of forces and balance of moments for every sub-body. We then demonstrate that the converse holds only for first-gradient continua. More precisely, for higher-gradient continua, a postulation scheme based \textit{solely} on balance of forces and moments does not uniquely determine equilibrium configurations, and a formulation based on the Principle of Virtual Work is therefore unavoidable.

In Section 4, we revisit Noll's argument that the surface contact force density depends only on the unit normal to the oriented surface over which the resultant interaction is evaluated. We identify precisely where implicit assumptions, namely the absence of edge and wedge contact forces and the boundedness of the surface density, enter the proof, and we show that these hypotheses fail for general higher-gradient continua. Consequently, the surface density may depend nontrivially on curvature, notwithstanding Noll's comment regarding the scope of his theorem.

In our concluding Section 5, we discuss the structural conditions under which Noll's theorem extends to higher-gradient continua, emphasizing the role of wedge and edge interactions and outlining directions for further investigation.

\section{Preliminaries}

\subsection{The Principle of Virtual Work}
We formulate the Principle of Virtual Work following the presentation of Paul Germain; for a more detailed development in its modern functional analytic form, we refer to his foundational works \cite{germain2020memocs, germain1973microstructure}. We consider a deformable body whose current configuration at a fixed time is given by the closure of a bounded domain $\mathcal B_t$ in three-dimensional Euclidean space. A sufficiently regular subset of $\mathcal B_t$ will be denoted by $\mathcal P_t$ and referred to as a sub-body of the body\footnote{More precisely, we will assume throughout that $\mathcal B_t$ and $\cl P_t$ are are the closures of domains in three-dimensional Euclidean space with boundaries given by piecewise regular surfaces; see, e.g., \cite{dellIsola2012contact}.}. Since time will be held fixed throughout our discussion, we omit the explicit dependence on $t$ for all sets and fields.
 
The assumptions of the Principle of Virtual Work can be split into the
following parts:
\begin{enumerate}
\item Internal interactions in a sub-body $\cl P \subseteq \cl B$ are modeled by a linear functional acting on a topological vector space $\mathcal A({\mathcal P})$ of admissible virtual displacements.
\item For each sub-body $\cl P \subseteq \cl B$, smooth functions are contained in the space of admissible virtual displacements, $C^{\infty}(\cl P)\subseteq \cl A(\cl P)$, and this inclusion is continuous.
\item The internal work associated with a sub-body $\mathcal P$ is modeled by a linear and continuous functional acting on $\mathcal A(\mathcal P)$, i.e., an element of the topological dual $\cl A(\cl P)'$. Consequently, the internal work functional may be identified with an element of $\cl E'(\cl P)$, a distribution in the sense of Schwartz supported on $\cl P$\footnote{This follows from (2) and the implication 
\begin{align}	
C^\infty(\cl P) \subseteq \cl A(\cl P) \implies \cl A(\cl P)' \subseteq C^\infty(\cl P)' = \cl E'(\cl P).
\end{align}}. We denote its action on an admissible virtual displacement $\boldsymbol v$ by $W^{int}(\mathcal P, \boldsymbol v)$. 
\item If $\cl P \subseteq \tilde{\cl P}$ are sub-bodies, then the restriction of $W^{int}(\cl P,\cdot)$ to admissible virtual displacements supported in $\tilde{\cl P}$ is equal to $W^{int}(\tilde{\cl P},\cdot)$.
\item The internal work functional for any sub-body $\cl P \subseteq \cl B$ applied to any \textit{rigidifying} virtual displacement vanishes; this property is often called objectivity of internal work\footnote{In the relevant literature, admissible rigidifying virtual displacements are defined as those virtual displacements for which the distance between every pair of material particles is preserved. Such displacements are not restricted to rigid bodies, but may also be realized as particular virtual motions of deformable bodies. The term rigidifying is therefore appropriate, as it emphasizes that the virtual displacement temporarily imposes a rigid behavior on the body under consideration.}.
\item External interactions acting on a sub-body $\mathcal P$ are modeled by a linear and continuous functional defined on the same space of admissible virtual displacements. This functional is referred to as the external work functional and is denoted by $W^{ext}(\mathcal P, \cdot)$. In general, the external work functional does not satisfy objectivity.

\item For every admissible virtual displacement $\bs v$ and sub-body $\cl P \subseteq \cl B$, we have 
\begin{equation}
	W^{int}(\cl P,\bs v)=W^{ext}(\cl P,\bs v). \label{eq:PVW}
\end{equation}
\end{enumerate}

We comment that a persistent source of misunderstanding in the literature concerns the status of the internal work functional within approaches based exclusively on balance laws. Critics have claimed that the Principle of Virtual Work is applicable only to linear constitutive relations, but this assertion rests on a fundamental conceptual confusion. The Principle of Virtual Work may be understood as a natural extension of the Principle of Minimum Energy to a broader class of mechanical systems. The Principle of Minimum Energy states that equilibrium configurations correspond to minimizers of the total energy functional. Euler and Lagrange established how to compute the first variation of this functional in order to enforce a stationarity condition. In modern terms, this procedure corresponds to evaluating the Fréchet derivative of the energy functional and requiring that it vanishes for all admissible infinitesimal variations about an equilibrium configuration. Crucially, even when the energy functional is non quadratic and leads to nonlinear equilibrium equations, its Fréchet derivative remains a linear functional of the admissible variations, and this linearity underlies the general validity of the Principle of Virtual Work.

The origin of the concept of virtual displacement is most naturally understood from the above variational perspective in which they correspond to small variations of the equilibrium configuration. We remark that both Lagrange and Piola devoted considerable effort to justifying the formal identity
	$
	\nabla^{i}\delta \boldsymbol u = \delta \nabla^{i} \boldsymbol u.
	$
	This identity, while correct, requires careful interpretation. In the original formulations, the symbol $\delta$ denotes a small variation measured in a norm that was not explicitly specified by either Lagrange or Piola. The commutation of the variation operator $\delta$ with the differential operator $\nabla^{i}$ therefore implicitly assumes that not only the field itself, but also its derivatives up to order $i$, are small in the relevant sense. In modern functional analytic terms, this requirement simply amounts to assuming that the internal work functional is defined on an appropriate Sobolev space, a setting in which such commutation properties are mathematically justified.

\subsection{Constraints on external work functionals}

As a direct consequence of \eqref{eq:PVW}, once an internal work functional has been specified for a deformable body, the class of admissible external work functionals is determined. This conclusion has sometimes been regarded as difficult to understand; we briefly present the argument originally articulated by Piola in modern terminology.

Consider, for example, a compressible Eulerian fluid whose deformation energy depends on the spatial mass density $\rho$. In this case, the internal work functional has a highly specific form, since it is given by the Fréchet derivative of the corresponding deformation energy functional. It then follows that the external work functional for a sub-body $\cl P$ can only be represented as the sum of a Lebesgue volume integral of the form $\int_{\mathcal P} \boldsymbol f \cdot \boldsymbol v \, dv$, and a surface integral over the boundary of the sub-body of the form $\int_{\partial \cl P} \boldsymbol t \cdot \boldsymbol v \, da$. Here the traction is necessarily given by $\boldsymbol t = - p \boldsymbol n$, where $p$ is a scalar field and $\boldsymbol n$ denotes the outward unit normal to the boundary. In particular, this structure makes clear that shear tractions cannot act on the boundary of a compressible Eulerian fluid.

It should therefore be expected that, once the structure of the internal work functional has been specified, the class of admissible external work functionals applicable to the body is correspondingly restricted. In particular, if the internal work functional is of order $n$, then the external work functional cannot be of higher order than $n$, although it may be of lower order.

\section{The Principle of Virtual Work and balance laws}

In this section, we demonstrate that while the Principle of Virtual Work implies balance of forces and moments for $n$th-gradient continua, the converse holds only for first-gradient continua, so that balance laws alone are insufficient to characterize equilibrium in higher-gradient theories.

\subsection{Piola's theorem in the distributional framework}

As recognized by Truesdell, see \cite{truesdell_tupin1960,Truesdell1991FirstCourseRCMVol1}, and demonstrated in \cite{dellIsola2014Piola1, dellIsola2014Piola2}, Gabrio Piola proved that for every $n$th-gradient continuum, a necessary condition for equilibrium is that the balance of forces and the balance of moments hold for every sub-body $\cl P \subseteq \cl B$. The argument we present below follows Piola's reasoning in substance, but is formulated using vector algebra and our functional analytic framework.

By Section 1.1 and Section 2.1, the external work functional associated to a sub-body $\cl P$ contained in an $n$th-gradient continuum $\cl B$ has the form:
\[
W^{ext}(\cl P,\bs v)=\sum_{i=0}^{n}\int_{\cl P} \nabla^i \bs v \cdot d\bs T_{i, \cl P}^{ext}, \quad \bs v \in \cl A(\cl P). 
\]
The external work functional must vanish when $\bs v$ is a rigidifying virtual displacement,
\[
\bs v(\bs x)=\bs v_{0}+\bs \omega\times(\bs x-\bs x_{0})=\bs v_{0}+\bs W(\bs x-\bs x_{0}), \quad \bs x \in \cl B, 
\]
by \eqref{eq:PVW} and objectivity of the internal work functional. Above, $\bs v_{0}$ is the displacement of the material point placed at $\bs x_{0}$
and $\bs W$ is the skew symmetric tensor with axial vector $\bs \omega$. Therefore, for every $\bs v_{0}$ and $\bs \omega$, we must have that 
\begin{align}
W^{ext}(\cl P,\bs v_{0}+\bs W(\bs x-\bs x_{0}))=\int_{\cl P } \bigl [\bs v_{0}+\bs W(\bs x-\bs x_{0})\big] \cdot d\bs T_{0, \cl P}^{ext} +\int_{\cl P} \bs W \cdot d\bs T_{1, \cl P}^{ext} = 0. \label{eq:extpower0}
\end{align}

By choosing $\bs W = \bs 0$ one gets for all $\bs v_0 \in \mathbb R^3$,
\begin{equation}
	W^{ext}(\cl P,\bs v_{0})=\bs v_0 \cdot \Bigl (\int_{\cl P}d\bs T_{0, \cl P}^{ext} \Bigr )=0.\label{eq:Vo}
\end{equation}
Introducing the resultant of external forces
\[
\bs f^{ext}(\cl P):=\int_{\cl P}d\bs T_{0, \cl P}^{ext}
\]
we conclude that equation \eqref{eq:Vo} implies the \textit{balance of forces} equation
\begin{align}
\bs f^{ext}(\cl P) = \bs 0. \label{eq:balanceforces}
\end{align}
In the variationally inspired mechanics literature, this relation is commonly referred to as the \textit{equation of equilibrium under translation}.

From \eqref{eq:extpower0} and \eqref{eq:Vo}, we conclude that for every skew tensor $\bs W$, 
\begin{align}
	\bs W \cdot \Bigl ( \int_{\cl P}
	d\bs T_{0,\cl P}^{ext} \otimes (\bs x - \bs x_0) + \int_{\cl P} d\bs T_{1, \cl P}^{ext}
	\Bigr ) = 0,
\end{align}
and thus, we have the vanishing of the skew part: 
\begin{align}
	\mathrm{skw}	\Bigl ( \int_{\cl P}
	d\bs T_{0, \cl P}^{ext} \otimes (\bs x - \bs x_0) + \int_{\cl P} d\bs T_{1, \cl P}^{ext}
	\Bigr ) = \bs 0. \label{eq:mo}
\end{align}
Introducing the resultant moment of external forces
\begin{align}
	\bs m^{ext}(\cl P) := \int_{\cl P} (\bs x - \bs x_0) \times d\bs T_{0, \cl P}^{ext}, 
\end{align}
we see that \eqref{eq:mo} is equivalent to the \textit{balance of moments} equation 
\begin{align}
	\bs m^{ext}(\cl P) + 2 \int_{\cl P} \mathrm{axl}( \mathrm{skw} \, d \bs T_{1, \cl P}^{ext}) = \bs 0. \label{eq:balancemoments}
\end{align}
The second term on the left-hand side of \eqref{eq:balancemoments} represents a resultant couple applied to $\cl P$.

Together, \eqref{eq:balanceforces} and \eqref{eq:balancemoments} constitute Piola's theorem for $n$th-gradient continua, i.e., the Principle of Virtual Work implies balance of forces and moments for every sub-body of $\cl B$.

\subsection{Indeterminacy of balance laws in higher-gradient continua}

It is well known that for first-gradient continua satisfying $d \boldsymbol T_{1,\cdot}^{\mathrm{ext}} = \boldsymbol 0$, the converse of Piola's theorem holds, so that the Principle of Virtual Work is equivalent to the postulation of balance of forces and balance of moments, see for example \cite{truesdell_noll2004}. We now show that a postulation scheme based solely on the balance laws \eqref{eq:balanceforces} and \eqref{eq:balancemoments} does not uniquely characterize equilibrium configurations for linear elastic second-gradient continua. In particular, these two balance laws are insufficient to determine the response of higher-gradient continua. A determinate theory therefore requires a formulation grounded in the more fundamental Principle of Virtual Work.

We consider a homogenous and isotropic \textit{linear dilatation strain-gradient} elastic body with stored energy
\begin{gather}
	W(\bs u) = \frac{1}{2}\lambda (\mathrm{tr}\, \bs \epsilon(\bs u))^2 + \mu |\bs \epsilon(\bs u)|^2 + \frac{\alpha}{2}|\nabla \mathrm{tr}\, \bs \epsilon(\bs u)|^2, \quad \bs \epsilon(\bs u) = \frac{1}{2}(\nabla \bs u + \nabla \bs u^T), \\
\lambda + \frac{2}{3}\mu > 0, \quad \mu > 0, \quad \alpha > 0,
\end{gather}
where $\bs u : \cl B \rightarrow \mathbb R^3$ is the displacement field and $\cl B$ is now the reference configuration \cite{EremeyevCazzaniDellIsola2021, EremeyevLurieSolyaevDellIsola2020}. In what follows, we assume that $\cl B$ and sub-bodies are $C^1$ domains\footnote{A conceptual framework demonstrating that linearized elasticity can be formulated in a frame-invariant manner, and therefore falls within the scope of Piola's theorem, is developed in \cite{Steigmann2007FrameInvariance}.}. The internal work functional associated to a static displacement $\bs u$ is defined to be the Frech\'et derivative of the stored energy functional,
\begin{align}
	W^{int}(\cl P, \bs v) := \int_{\cl P} \Bigl [\lambda (\mathrm{div}\, \bs u)(\mathrm{div}\, \bs v) + 2\mu \bs \epsilon(\bs u) \cdot \bs \epsilon(\bs v) + \alpha \nabla \mathrm{div}\, \bs u \cdot \nabla \mathrm{div}\, \bs v \Bigr ] dv, \quad \bs v \in \cl A(\cl P),
\end{align}
and the class of admissible external work functions (see the discussion in Section 2.2) takes the form 
\begin{align}
	W^{ext}(\cl P, \bs v) := \int_{\cl P} \bs f \cdot \bs v \, dv + \int_{\partial \cl P} \Bigl (
	\bs t \cdot \bs v + c \frac{\partial \bs v}{\partial n} \cdot \bs n
	\Bigr )da, \quad \bs v \in \cl A(\cl P), 
\end{align}
where $\bs f$ is an external body force, $\bs t$ is a traction, $c \bs n$ is a surface double force, and $\bs n$ is the unit outward normal (see Section 3 of \cite{EremeyevLurieSolyaevDellIsola2020}). An equilibrium configuration corresponds to a displacement $\boldsymbol u$ satisfying \eqref{eq:PVW}. This condition is equivalent to requiring that $\boldsymbol u$ solve a certain fourth order partial differential equation in $\mathcal P$, together with explicit boundary relations linking $\boldsymbol u$ to $\boldsymbol t$ and $c \boldsymbol n$ on $\partial \mathcal P$ (see Section 3 of \cite{EremeyevLurieSolyaevDellIsola2020} for precise details).
 
For an equilibrium configuration, it is straightforward to see that \eqref{eq:balanceforces} and \eqref{eq:balancemoments} are given, respectively, by 
\begin{align}
\int_{\cl P} \bs f \, dv + \int_{\partial \cl P} \bs t \, da &= \bs 0, \label{eq:balanceforcesdil}\\
\int_{\cl P} (\bs x - \bs x_0) \times \bs f \, dv + \int_{\partial \cl P} (\bs x - \bs x_0) \times \bs t \, da &= \bs 0, \label{eq:balancemomdil}
\end{align}
for every sub-body $\cl P \subseteq \cl B$. The standard formulation of the mixed boundary value problem, based solely on the balance laws \eqref{eq:balanceforcesdil} and \eqref{eq:balancemomdil}, suggests that prescribing the body force $\boldsymbol f$ in $\mathcal B$, together with the traction $\boldsymbol t$ and the displacement behavior on complementary subsets $\partial \cl B_0$ and $\partial \cl B_1$ of $\partial \mathcal B$, uniquely determines the corresponding equilibrium configuration; see, e.g., \cite{Gurtin1972LinearElasticity} for classical first-gradient linearized elasticity\footnote{In this case, $\cl A(\cl B)$ would be contained in the set of smooth functions on $\cl B$ that are $\bs 0$ on $\partial \cl B_1$, and $\cl A(\cl P) = C^\infty(\cl P)$ for $\cl P$ contained in the interior of $\cl B$.}.

This conclusion is, however, \textit{incorrect}. As shown in Section 4 of \cite{EremeyevLurieSolyaevDellIsola2020}, once $\boldsymbol f$, $\partial \cl B_0$, $\partial \cl B_1$, and $\boldsymbol t$ are fixed, there exist \textit{infinitely} many displacements satisfying the balance equations \eqref{eq:balanceforcesdil} and \eqref{eq:balancemomdil}, the prescribed traction $\boldsymbol t$ on $\partial \cl B_0$, and the condition $(\boldsymbol u|_{\partial \cl B_0}, \partial_n \boldsymbol u \cdot \bs n|_{\partial B_1}) = (\boldsymbol 0, 0)$. Each such displacement corresponds to a distinct choice of double force density $$c : \partial \cl B_1 \to \mathbb R$$ which \textit{does not appear in the balance laws}. It follows that a postulation scheme based solely on the two basic balance laws is insufficient to determine the response of linearly elastic dilatational strain gradient continua, and more generally of higher-gradient continua, and a formulation grounded in the Principle of Virtual Work is essential.

\section{The scope of Noll's theorem}

In this section, we examine Noll's theorem and proof and show that its conclusion relies essentially on assumptions that fail for higher-gradient continua.

\subsection{Preliminary set-up and hypotheses} In this section, we consider a family of vector-valued measures $\{ d\bs T^{ext}_{0, \cl P}\}$ indexed by the sub-bodies of $\cl B$ such that each measure consists of four parts:
\begin{enumerate}
	\item a measure $\bs b(\cl P; \bs x) dv$ that is absolutely continuous with respect to Lebesgue measure $dv$ on $\cl P$,
	\item a measure supported on the wedges of $\partial \cl P$,
	\item a measure supported on the edges of $\partial \cl P$, 
	\item a measure supported on the faces, $\cl S_{\partial \cl P}$, of the boundary of $\cl P$, that is absolutely continuous with respect to surface measure.  
\end{enumerate}
 We denote by $\boldsymbol B(\cl P),$ $\boldsymbol W(\cl P)$, $\bs E(\cl P)$ and $\boldsymbol S(\mathcal P)$ the measures of $\mathcal P$ associated with the first through fourth components of $d\boldsymbol T^{ext}_{0,\mathcal P}$, respectively. We refer to these vectors as the \textit{resultant body force}, \textit{resultant wedge contact force}, \textit{resultant edge contact force}, and \textit{resultant surface contact force} exerted on $\cl P$, respectively. 
 
We first assume the existence of a constant $C > 0$ such that, for every sub-body $\mathcal P$ and every $\boldsymbol x \in \mathcal P$,
\begin{align}
	|\boldsymbol b(\mathcal P; \boldsymbol x)| \leq C, \label{eq:bodybounded}
\end{align}
a uniform boundedness condition that is natural for standard body force densities $\boldsymbol b$ that do not depend on the choice of sub-body. For the surface contact force, we assume the existence of a vector-valued function $\boldsymbol s$ defined on pairs $(\mathcal S; \boldsymbol x)$, where $\mathcal S$ is a smooth orientable surface contained in $\mathcal B$, such that for each oriented surface $\mathcal S$, the map $\boldsymbol x \mapsto \boldsymbol s(\mathcal S; \boldsymbol x)$ is integrable on $\mathcal S$ with respect to surface measure. The resultant surface contact force acting on a sub-body $\mathcal P$ is postulated to be given by
\begin{align}
	\boldsymbol S(\mathcal P)
	=
	\sum_{\mathcal S \in \mathcal S_{\partial \mathcal P}}
	\int_{\mathcal S} \boldsymbol s(\mathcal S; \boldsymbol x)\, da. \label{eq:surfcont}
\end{align}
Finally, we assume locality with respect to the surface, in the sense that if $\mathcal S' \subseteq \mathcal S$ is an oriented sub-surface, then
\begin{align}
	\boldsymbol s(\mathcal S'; \boldsymbol x)
	=
	\boldsymbol s(\mathcal S; \boldsymbol x),
	\qquad
	\boldsymbol x \in \mathcal S'. \label{eq:locality}
\end{align}

Similar to Section 3, we define the \textit{resultant of external forces} on $\cl P$ by 
\begin{align}
\bs f^{ext}(\cl P) := \int_{\cl P} d\bs T_{0, \cl P}^{ext} = \bs B(\cl P) + \bs W(\cl P) + \bs E(\cl P) + \bs S(\cl P), \label{eq:extforces}
\end{align}
and we assume balance of forces: for every sub-body $\cl P$, 
\begin{align}
	\bs f^{ext}(\cl P) = \bs 0. \label{eq:balextforces}
\end{align}

\subsection{Noll's additional assumptions and theorem}

In addition to the preliminary assumptions introduced in the previous section, Noll's theorem relies on the following fundamental hypotheses, see \cite{Noll1974FoundationsReprint, truesdell1991PrefaceAntman}:
\begin{itemize}
	\item The body does not support wedge or edge interactions; that is, for every sub-body $\mathcal P$,
	\begin{align}
		\boldsymbol W(\mathcal P) = \boldsymbol E(\mathcal P) = \boldsymbol 0.
		\label{eq:nollass1}
	\end{align}
	\item The surface density $\boldsymbol s$ is uniformly bounded on its domain of oriented surfaces contained in $\mathcal B$; specifically, there exists a constant $C > 0$ such that for every oriented surface $\mathcal S \subset \mathcal B$ and every $\boldsymbol y \in \mathcal S$,
	\begin{align}
		|\boldsymbol s(\mathcal S; \boldsymbol y)| \leq C.
		\label{eq:Nollassumpt}
	\end{align}
\end{itemize}
Under the assumptions of Section~4.1 together with \eqref{eq:nollass1} and \eqref{eq:Nollassumpt}, Noll's theorem, in a form essentially identical to Theorem~IV of \cite{Noll1974FoundationsReprint} (see also Section~III.3 of \cite{truesdell1991PrefaceAntman}), may be stated as follows.
\begin{thm}\label{t:1}
	There is a vector-valued function $\boldsymbol s(\boldsymbol x, \boldsymbol n)$, where $\boldsymbol x \in \cl B$ and where $\boldsymbol n$ is a unit vector, such that 
	\begin{align}
		\boldsymbol s(\mathcal S; \boldsymbol x) = \boldsymbol s(\boldsymbol x, \boldsymbol n), 
	\end{align}
	whenever the oriented surface $\mathcal S \subset \cl B$ has the unit normal $\boldsymbol n$ at $\boldsymbol x$ directed towards the positive side of $\mathcal S$. 
\end{thm}

\subsection{Noll's argument}

\begin{figure}[t]
	\centering
	\includegraphics[width=.6\linewidth]{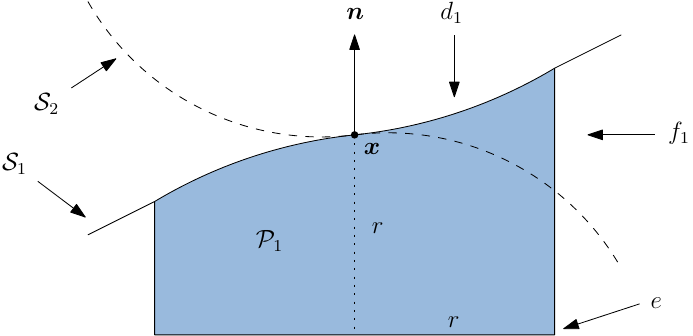}
	\caption{}
	\label{f:1}
\end{figure}

We pinpoint the stages in Noll's proof of Theorem \ref{t:1} \cite{Noll1974FoundationsReprint} at which the additional assumptions \eqref{eq:nollass1} and \eqref{eq:Nollassumpt} are utilized for points $\boldsymbol x$ lying in the interior of $\cl B$. Let $\mathcal S_1$ and $\mathcal S_2$ be two surfaces in the body's current configuration $\cl B$ tangent to each other at $\boldsymbol x$. We assume that $\boldsymbol n$ is oriented toward the positive side of both $\mathcal S_1$ and $\mathcal S_2$. Consider the region $\mathcal P_1$ bounded by a piece $d_1$ of $\mathcal S_1$, a piece $f_1$ of a circular cylinder of radius $r$ whose axis is $\boldsymbol n$, and by a plane perpendicular to $\boldsymbol n$ at a distance $r$ from $\boldsymbol x$ (see Figure \ref{f:1}). The region $\mathcal P_2$ is defined in a similar manner. We denote the common boundary of $\mathcal P_1$ and $\mathcal P_2$ on the cylinder and the plane by $e$. The boundaries of $\mathcal P_1$ and $\mathcal P_2$ then have decompositions into separate faces
\begin{align}
	\partial \mathcal P_1 = d_1 \cup e \cup f_1, \quad \partial \mathcal P_2 = d_2 \cup e \cup f_2. \label{eq:boundarydec}
\end{align}
Since $\boldsymbol x$ is in the interior of $\cl B$, for all $r$ sufficiently small, $\mathcal P_1$ and $\mathcal P_2$ will be sub-bodies of $\cl B$. We denote the surface area of a surface $\mathcal S$ by $A(\mathcal S)$. 
It follows that for $i = 1,2$, 
\begin{align}
	A(d_i) = \pi r^2 + o(r^2), \quad A(f_i) = o(r^2). \label{eq:area} 
\end{align}

For $i = 1,2$, by \eqref{eq:surfcont} and \eqref{eq:locality}, the resultant surface contact force exerted on $\mathcal P_i$ is given by 
\begin{align}
	\boldsymbol S(\mathcal P_i) = \int_{d_i} \boldsymbol s(\mathcal S_i) \, dA + \int_{f_i} \boldsymbol s(f_i) \, dA + \int_e \boldsymbol s(e)\, dA.  
\end{align}
Subtracting the two equations yields 
\begin{align}
	\begin{split}
		\int_{d_1} &\boldsymbol s(\mathcal S_1) \, dA - \int_{d_2} \boldsymbol s(\mathcal S_2) \, dA \\ &= \boldsymbol S( \mathcal P_1) - \boldsymbol S(\mathcal P_2) - \int_{f_1} \boldsymbol s(f_1) \, dA + \int_{f_2} \boldsymbol s(f_2) \, dA. \label{eq:difference}
	\end{split}
\end{align}
Assuming that $\boldsymbol x$ is in the Lebesgue set of $\boldsymbol s(\mathcal S_i; \cdot)$ for $i = 1,2$, it follows via \eqref{eq:area} and the standard localization theorem that 
\begin{align}
	\lim_{r \rightarrow 0} \frac{1}{\pi r^2} \int_{d_i} \boldsymbol s(\mathcal S_i) \, dA = \lim_{r \rightarrow 0} \frac{1}{A(d_i)} \int_{d_i} \boldsymbol s(\mathcal S_i) \, dA =  \boldsymbol s(\mathcal S_i; \boldsymbol x).  \label{eq:localization}
\end{align}

It is at this stage of the argument that Noll's additional assumptions \eqref{eq:nollass1} and \eqref{eq:Nollassumpt} play \textit{crucial} roles in showing that the right-hand side of \eqref{eq:difference} vanishes as $r \rightarrow 0$, thereby completing the proof. In particular, using the \textit{boundedness assumption} \eqref{eq:Nollassumpt} together with the area estimate \eqref{eq:area}, one obtains for
	for $i = 1, 2$, 
	\begin{align}
		\int_{f_i} \boldsymbol s(f_i) \, dA = o(r^2). \label{eq:fibound}
	\end{align}
	By \eqref{eq:area}, it then follows that for $i = 1,2$, 
	\begin{align}
		\lim_{r \rightarrow 0} \frac{1}{\pi r^2} \int_{f_i} \boldsymbol s(f_i) \, dA = \lim_{r \rightarrow 0} \frac{o(r^2)}{\pi r^2} = 0. \label{eq:fvanishing}
	\end{align}
Then \eqref{eq:difference}, \eqref{eq:fvanishing}, and \eqref{eq:localization} imply that 
\begin{align}
	\boldsymbol s(\mathcal S_1; \boldsymbol x) - \boldsymbol s(\mathcal S_2; \boldsymbol x) = \lim_{r \rightarrow 0} \frac{1}{\pi r^2} [\boldsymbol S(\mathcal P_1) - \boldsymbol S(\mathcal P_2)] \label{eq:differenceatx}
\end{align}
Combining \eqref{eq:extforces}, \eqref{eq:balextforces}, \eqref{eq:bodybounded}, the assumption of \textit{absence of wedge and edge forces} \eqref{eq:nollass1}, and the area estimate \eqref{eq:area}, we deduce that for $i = 1,2$,
\begin{align}
	\boldsymbol S(\mathcal P_i) = O\bigl(V(\mathcal P_i)\bigr) = o(r^2).
	\label{eq:Nollbound}
\end{align}
It then follows from \eqref{eq:differenceatx} and \eqref{eq:Nollbound} that 
\begin{align}
	\boldsymbol s(\mathcal S_1; \boldsymbol x) - \boldsymbol s(\mathcal S_2; \boldsymbol x) = \boldsymbol 0,
\end{align}
proving Theorem \ref{t:1} for the case that $\boldsymbol x$ is in the interior of $\cl B$. 

\subsection{Non-applicability of Noll's theorem to higher-gradient continua}
The conclusion of Noll's theorem, Theorem~\ref{t:1}, together with his accompanying remark quoted in the Introduction, appears to conflict with what is well known for $n$th-gradient materials. In particular, for general second-gradient materials, which form a subclass of $n$th-gradient materials with $n \geq 2$, the surface contact force density $\boldsymbol s$ may depend nontrivially on the normal curvature tensor of the oriented surface $\mathcal S$, see \cite{Toupin1964CoupleStress, germain2020memocs}. This apparent contradiction is resolved by recalling that Theorem~\ref{t:1} rests on additional hypotheses, namely the absence of wedge or edge interactions \eqref{eq:nollass1} and the boundedness assumption on the surface contact force density \eqref{eq:Nollassumpt}. These assumptions do not hold for general second-gradient continua. Consequently, it is the restrictive nature of the hypotheses, rather than the theorem's conclusion, that excludes the more general surface stresses characteristic of higher-gradient materials.
  
To see that the assumption \eqref{eq:Nollassumpt} does not hold for general second-gradient continua, consider a second-gradient elastic material with stored energy
\begin{align}
	W = W_0(\boldsymbol E) + \frac{\alpha}{2}|\nabla \boldsymbol F|^2 
\end{align}
where $\boldsymbol \chi(\cdot) : \cl B_r \rightarrow \cl B$ is the deformation at time $t$, $\cl B_r$ is a reference configuration, $\boldsymbol F = \nabla \boldsymbol \chi$ is the deformation gradient, and $\boldsymbol E = \frac{1}{2}(\boldsymbol F^t \boldsymbol F - \boldsymbol I)$ is the Lagrange strain. Suppose without loss of generality that $(0,0,0) \in \cl B$, and the deformation is given by 
\begin{align}
	\boldsymbol x = \boldsymbol \chi(\boldsymbol X) = (\lambda (e^{X_1/\lambda} - 1), X_2, X_3), 
\end{align}
with $\lambda > 0$ a fixed length scale. Then 
\begin{gather}
	\boldsymbol \chi(0,0,0) = (0,0,0), \quad \boldsymbol F(0,0,0) = \boldsymbol I, \label{eq:def1} \\
	\nabla \boldsymbol F(\boldsymbol X) = \lambda^{-1} e^{X_1/\lambda} \boldsymbol e_1 \otimes \boldsymbol e_1 \otimes \boldsymbol e_1. \label{eq:def2}
\end{gather}
Let $\{ \mathcal S_r \}_{r > 0}$ denote the family of \textit{shrinking} right circular cylinders
\begin{align}
	\mathcal S_r = \bigl\{ (x_1, x_2, x_3) \,\big|\, (x_1 + r)^2 + x_2^2 = r^2,\; x_3 \in [-r/2, r/2] \bigr\},
\end{align}
each of equal height along the $x_3$-axis and radius $r \ll \mathrm{diam}(\cl B)$, such that their lateral surfaces pass through the origin and share the common normal $\boldsymbol e_1$ at $(0,0,0)$. Then, using \eqref{eq:def1}, \eqref{eq:def2} and the formulae given on page 102 of \cite{Toupin1964CoupleStress}, one can show that the supporting surface tractions satisfy 
\begin{align}
	\boldsymbol s(\mathcal S_r; (0,0,0)) = \frac{\alpha}{\lambda r} (\boldsymbol e_1 + \boldsymbol o(1) ) \rightarrow \infty, \quad \mbox{as } r \rightarrow 0. \label{eq:sblowup}
\end{align}
Thus, \eqref{eq:Nollassumpt} does not hold for general second-gradient materials. We emphasize that the blow up behavior described in \eqref{eq:sblowup} renders the crucial identity \eqref{eq:fibound} unattainable in Noll's proof, since the surfaces $f_i$ lie on the lateral sides of shrinking cylinders of the type described above.

Moreover, it is well established that general higher-gradient continua possess edge and wedge contact interactions and therefore do not satisfy Noll's additional assumption \eqref{eq:nollass1}, see \cite{DellIsolaSeppecher1997Edge, dellIsola2012contact}. In particular, the results of \cite{EremeyevDellIsola2022WeakSolutions} show that, given a sub-body $\mathcal P$ with edges contained in a $C^1$ linear dilatational strain gradient elastic body $\mathcal B$, there exist infinitely many equilibrium configurations of $\mathcal B$ that support edge contact forces along $\mathcal P$. Each such configuration arises from integrating an edge force density along the edges and is parametrized by the jump of the double force $c$ across those edges. We refer to Section~3.2 for the discussion of $c$ and to \cite{EremeyevDellIsola2022WeakSolutions} for the corresponding existence results. These results demonstrate that edge forces are not exceptional phenomena but intrinsic features of the mechanical response of higher-gradient continua; consequently, Noll's additional assumptions, and therefore Theorem~\ref{t:1}, are not applicable in this setting.

\section{Conclusion}

By formulating internal and external interactions within the distributional framework emphasized by Paul Germain, we have revisited foundational questions in Continuum Mechanics from a functional analytic standpoint. In the spirit of Piola, we have argued that the Principle of Virtual Work provides the appropriate structural foundation for higher-gradient continuum theories. Although balance of forces and moments remain necessary conditions for equilibrium, they are no longer sufficient beyond the first-gradient setting. In addition, our reexamination of Noll's theorem reveals that its conclusion relies on additional assumptions that are not satisfied in general higher-gradient continua. Thus, curvature-dependent surface interactions in such materials do not conflict with Noll's result, but fall outside the range of its applicability.

A natural direction for further investigation is the classification of higher-gradient materials for which the conclusion of Noll's theorem remains valid. It is therefore appropriate to begin with second-gradient continua, which do not admit wedge interactions. Results from \cite{DellIsolaSeppecher1997Edge} establish that, for second-gradient continua equipped with natural forms of internal and external work functionals from \cite{dellIsola2012contact} and satisfying the Principle of Virtual Work, the absence of edge contact interactions as defined in Section~4.1 suffices to recover the conclusion of Noll's theorem. Indeed, Theorem~9 of \cite{DellIsolaSeppecher1997Edge} shows that there exist a second-order tensor field $\bs T$ and a third-order tensor field $\bs C$ that encodes edge contact interactions such that
\begin{align}
	\bs s(\cl S; \bs x)
	=
	\bs T(\bs x)\bs n
	-
	\mathrm{div}_s\big((\bs C(\bs x) \bs n)(\bs I - \bs n \otimes \bs n)\big),
	\label{eq:sdellrep}
\end{align}
where $\bs n$ denotes the unit normal to $\cl S$ at $\bs x$ oriented in the positive direction and $\mathrm{div}_s$ denotes the surface divergence on $\cl S$. Theorems~7 and 8 of \cite{DellIsolaSeppecher1997Edge} imply that, if the body does not support edge contact interactions, then there exists a vector field $\bs c$ on $\cl B$ such that $\bs C = \bs c \otimes \bs I$. In this case, the second term on the right-hand side of \eqref{eq:sdellrep} vanishes, and the surface contact force density reduces to the classical form required by Noll's theorem. Whether the absence of edge interactions is not only sufficient but also \textit{necessary} for the validity of Noll's conclusion in second-gradient continua remains an open question that merits careful analysis.

More broadly, the structural representations of wedge, edge, and surface contact forces derived in \cite{dellIsola2012contact} suggest a systematic route for determining the admissible form of surface contact forces under the assumption that wedge and edge contact interactions are excluded. Advancing this program would work towards establishing the structural boundary that defines the true scope of Noll's theorem in higher-gradient continua.

\section*{Acknowledgements}

This paper was suggested by Professor Giampiero Del Piero to the second
author many years ago. Only with the efforts of the first author
could it be written. 

\bibliographystyle{abbrv}
\bibliography{biblioIvan}

\bigskip

\footnotesize

\noindent \textsc{Department of Mathematics, The University of North Carolina at Chapel Hill, USA}\\
\noindent \textit{E-mail address}: \texttt{crodrig@email.unc.edu}

\bigskip	

\noindent \textsc{Dipartimento Ingegneria Civile Edile Architettura e Ambientale, Università dell'Aquila,
	L'Aquila, Italy}\\
\textsc{CNRS Fellow, ENS Paris-Saclay, France} \\
\noindent \textit{E-mail address}: \texttt{francesco.dellisola@univaq.it}

\end{document}